# Low Temperature Combustion Synthesis of a Spinel NiCo$_2$O$_4$ Hole Transport Layer for Perovskite Photovoltaics

*Ioannis T. Papadas,[a] Apostolos Ioakeimidis,[a] Gerasimos S. Armatas[b] and Stelios A. Choulis[a]\**

[a] Molecular Electronics and Photonics Research Unit, Department of Mechanical Engineering and Materials Science and Engineering, Cyprus University of Technology, Limassol, Cyprus.

[b] Department of Materials Science and Technology, University of Crete, Heraklion 71003, Greece.

\*Corresponding Author: Prof. Stelios A. Choulis

E-mail: stelios.choulis@cut.ac.cy

**Abstract**

In the present study, we report the synthesis and characterization of a low-temperature solution-processable monodispersed nickel cobaltite (NiCo2O4) nanoparticles via a combustion synthesis using tartaric acid as fuel and demonstrate its performance as hole transport layer (HTL) for Perovskite Solar Cells (PVSCs). NiCo2O4 is a p-type semiconductor consisting of environmentally friendly, abundant elements and higher conductivity compared to NiO. We show that the combustion synthesis of spinel NiCo2O4 using tartaric acid as fuel can be used to control the NPs size and provide smooth, compact and homogeneous functional HTLs processed by blade coating. Study of PVSCs with different NiCo$_2$O$_4$ thickness as HTL reveal a difference on hole extraction efficiency and for 15 nm optimized thickness enhanced hole carrier collection is achieved. As a result, p-i-n structure of PVSCs with 15 nm NiCo$_2$O$_4$ HTLs showed reliable performance and power conversion efficiency values in the range of 15.5 % with negligible hysteresis.

## 1. Introduction

Over the last few years, a great deal of effort has been made to improve photovoltaic performance based on organic-inorganic lead halide perovskites, which has been reported to exhibit power conversion efficiencies (PCE) over 20 %.[1–5] The use of organic-inorganic lead halide perovskites has attracted intense interest due to extraordinary characteristics such as high light absorption,[6–9] enhanced charge transport properties, and direct band gap transition. For the fabrication of efficient perovskite solar cells, the so-called n-i-p architecture is the most common.[10] For the p-i-n-type perovskite solar cells, also called inverted architecture structure, poly(3,4-ethylenedioxythiophene): poly(styrenesulfonate) (PEDOT:PSS) is commonly used as a hole transport layer (HTL) and deposited first, followed by the perovskite layer and an n-type semiconductor film. PEDOT:PSS is an efficient and cost-effective HTL due to its facile processing, stable work function, and good electrical conductivity and transparency.[11–16] On the other hand, the hydroscopicity and inhomogeneous electrical properties, limit its application as a HTL.[17,18] Recently, p-type metal oxides and complexes, such as NiO, V$_2$O$_5$, CuO, CuSCN, CuPc and ZnPc [19–25] have been introduced into perovskite solar cells as a HTL. Inorganic p-type semiconductor materials have the advantages of improved hole selectivity and chemical stability, showing promising prospects as HTLs in perovskite solar cells.[26,27]

Up to now, sol–gel method is the most commonly used technique for the fabrication of the p-type metal oxides. However, in order to achieve the required crystallinity, temperatures above 400 °C are usually required. The need for high temperature is increasing fabrication cost and limits their potential use for printed electronic applications. [28] Thus there is a demand for the development of metal oxides using preparation methods that require lower temperatures.

Among many processes for the synthesis of nanomaterial compounds, combustion synthesis emerges as an efficient alternative approach. The combustion synthesis, in principles, can be defined as a redox (reduction/oxidation) or electron transfer process, in which the fuel is oxidized (increase of the oxidation state) and the oxidizer is reduced (reduce of the oxidation state) in an exothermic reaction.[29–32] Various types of combustion synthesis processes have been applied to obtain nanoparticles and they can be categorized according to the educts (gaseous, liquid or solid) and the process (e.g. combustion synthesis in the gas, solid or liquid phase, volume combustion synthesis, self-propagating high-temperature synthesis etc.).[33–35].

The combustion technique appears to be versatile and effective for the synthesis of high crystallinity solution-processed metal oxides thin films using low temperature.[36–39] Since it is an exothermic process, with a high heat release rate, the need for high temperatures is avoided and the production of high purity and homogeneous nanoparticles formation is simultaneously achieved.[40–42] For the production of metal oxides, liquid phase combustion synthesis has proven to be the most suitable, where usually metal salts (for instance nitrates) serve as oxidizers dissolved in saturated aqueous or alcoholic solutions in combination with organic fuels (e.g. urea, glycine, citric acid and others).[35,43,44] Upon heating gelation occurs and then combustion process starts resulting in the synthesis of the corresponding metal oxide.[33,45] The combustion synthesis of metal oxides exhibits great advantages comparing to other nanoparticle synthesis methods; namely, simple experimental setup, reduced number of post-processing steps, formation of nanoparticles without agglomeration, high purity of materials and precise control of particle's size and crystallinity by adjusting the processing parameters.[30,32,46–50] In general, the reaction mechanism of the combustion is affected by many factors such as the type of fuel, fuel-to-oxidizer ratio, ignition temperature and the H2O content of the precursor blend.[33,34,51,52]

Nickel cobaltite ($NiCo_2O_4$) is a p-type transparent conductive oxide (TCO) semiconductor consisting of abundant and environmentally friendly elements (Co, Ni), with a relatively wide optical band gap (~2.1 - 2.4 eV), deep-lying valence band (VB of 5.3 eV) that matches well with the VB of $CH_3NH_3PbI_3$ perovskite semiconductor and a much better conductivity than NiO and Co3O4 (at least two orders of magnitude higher).[53] These characteristics render NiCo2O4 one of the most promising candidates for electronic applications. $NiCo_2O_4$ adopts a cubic spinel structure in which all the Ni ions occupy the octahedral sites and the Co ions are distributed between the tetrahedral and octahedral sites. [53–55] It possesses high physical and chemical stability which is a necessity for high performance electronic devices. These attractive features make NiCo2O4 an appropriate candidate material for introduction as hole transport layer in PVSCs to achieve high-performance photovoltaic devices. NiCo2O4 derivatives have been used previously in many other applications such as anodic oxygen evolution,[56] inorganic and organic electrosynthesis,[57] development of supercapacitors[53,58] or infrared transparent conducting electrodes, sensors, optical limiters and switches,[59,60] but before this publication not for any type of solar cells.

Up to now various low-temperature synthetic routes such as hydrothermal, co-precipitation,[61,62] and thermal decomposition of the precursors such as hydroxide nitrates[63,64] and hydrazine carboxylate hydrates,[65] have been developed for the synthesis of NiCo2O4. Moreover, nanostructured aggregates of NiCo2O4 have been synthesized by employing heterometallic alkoxide precursor in the presence of a supramolecular liquid.[66] However, the production of high purity and monodispersed nanoparticles with the above mentioned synthesis approaches has not been completely achieved.

In this work, we present a one-step synthesis of low-temperature solution-processable nickel cobaltite (NiCo2O4) via combustion chemistry proposing for the first-time tartaric acid as a fuel and nitrate as an oxidizer agent. NiCo2O4 nanoparticles (NPs) with an average size of ~4 nm and narrow particle-size distribution were prepared using a cost-effective, low-temperature combustion synthesis method calcinated at 250 $^o$C for 1 hour. Those ultrafine nanoparticles enable the formation of compact, very smooth, high electrically conductive and relatively optically transparent NiCo2O4 films, which were utilized, for the first time, as HTLs in a solution processed p-i-n perovskite solar cell (PVSC). The effect of $NiCo_2O_4$-HTL thickness on PVSC characteristics is also investigated. A comparative study of devices incorporating different thickness of NiCo2O4-HTLs reveal a difference in hole extraction efficiency. The photo luminance spectroscopy measurements on perovskite films shown a reduced electron-hole pair recombination for the optimized 15 nm thick NiCo2O4-HTL. Additional electro-impedance spectroscopy and Mott-Schottky measurements on PVSC confirm the better hole extraction inducing an enhancement in the PVSC characteristics and negligible PCE hysteresis. The corresponding PVSC exhibits a high FF ≈ 80 % and a reliable performance with PCE of 15.5 %. The results demonstrate the great potentials of applying the low-temperature combustion synthesis for fabrication of highly reproducible and reliable metal oxide NPs which can be used for the formation of HTLs in variety of solution processed printed electronic devices.

## 2. Results and Discussion

Combustion synthesis has been applied recently for the low-temperature fabrication of metal oxide thin films.[67] In general, solution combustion synthesis has the advantage of rapidly producing homogeneous metal oxide materials with fine grain size, and most significantly at much lower temperature compared with the conventional solid-state reaction processes and co-precipitation methods. The structural and morphological characteristics of the resulting materials closely depend on the type and amount of chemicals (fuel, oxidizing agent) used in the synthesis.[33,45] Furthermore, the choice of the fuel regent for the combustion process has an essential role to avoid the formation of large clusters or/and large voids between the grains.[46] A fuel, i.e. the substance capable of acting as electrons acceptor, can significantly affect the properties of the final product, such as grain size, surface area, morphology, crystal phase, and degree and nature of particle agglomeration.[52,55]

In this work, tartaric acid was used as a fuel that is critically important to obtain uniform single-crystalline phase $NiCo_2O_4$ NPs. The advantage of using tartaric acid is related to the formation of heterometallic polynuclear complexes[68,69] due to the presence of its carboxylate and hydroxyl groups in a proper orientation, where the binding metal ions (i.e., $Ni^{2+}$ and $Co^{3+}$) come close together.[70] Ultimately, the formation of nickel cobaltite NPs is the consequence of decomposition of polynuclear complexes upon mild heating in the presence of concentrated $HNO_3$.[71]

**2.1. Synthesis and Characterization $NiCo_2O_4$ Nanoparticles**

It is well documented in the literature that thin metal oxide films can be obtained at temperatures lower than bulk-like powders via combustion synthesis because of the enhanced gas transport mechanisms and the easier out-diffusion of volatile products.[72] This means that thin films may decompose at low temperature and exhibit high sensitivity to any residual reactive gas present in the oven.[67,73]

For the combustion synthesis of $NiCo_2O_4$, metal precursors (Ni and Co nitrates) and tartaric acid (at a 1:2 molar ratio, respectively) were dissolved in 2-methoxyethanol containing a small amount of 0.24 mM $HNO_3$ (**Figure 1**a). The combustion reaction of the Ni(II)/Co(III)-tartaric complexes was monitored by differential scanning calorimetry (DSC) and thermogravimetric analysis (TGA), applying a heating rate of 10 °C/min in air. As shown in Figure 1b, the reaction exhibits an intense exothermic peak at ≈260 °C in the DSC curve, which coincides well with the abrupt mass loss (at ≈250 °C) observed in TGA curve. This implies that the formation of $NiCo_2O_4$ NPs via such combustion method can effectively proceed at a much lower temperature.

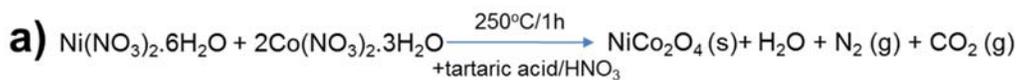

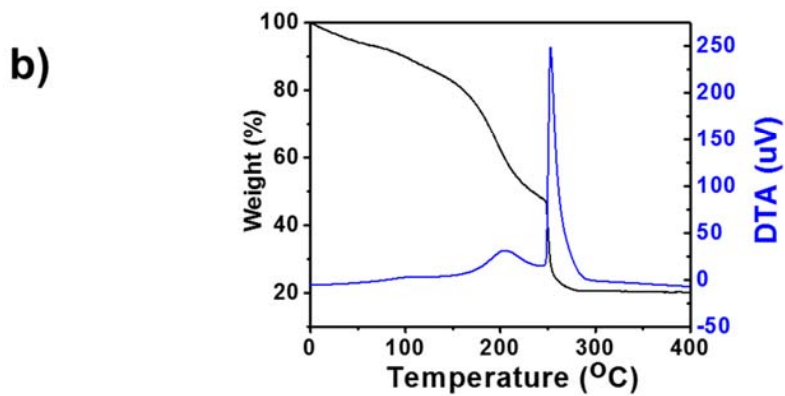

**Figure 1.** (a) Depiction of the synthetic route for spinel $NiCo_2O_4$, (b) TGA and DTA profiles of the as-prepared $NiCo_2O_4$ via combustion process.

In order to crystallize the as-synthesized films to spinel phase, 1 hour of heating was applied at different temperatures, i.e. 200, 250 and 300 °C, and the XRD results of the obtained materials are shown in **Figure 2**a. The XRD patterns of the materials calcined at 250 and 300 °C correspond to the spinel phase of $NiCo_2O_4$, although with a larger grain composition for the sample treated at 300 °C, as indicated by the narrow full width of half-maximum (FWHM) of XRD peaks. The angular position of the diffraction peaks matches well with standard XRD pattern of cubic spinel $NiCo_2O_4$ with JCPDS card no 20–0781. Notably, we did not observe any additional peaks arising from impure phases, indicating the single-crystalline nature of samples. The average grain of the $NiCo_2O_4$ NPs was estimated from the diffraction peak (220) by using the Scherrer's equation and was found to be ~3.5 nm for the sample annealed at 250 °C and ~5 nm for the sample annealed at 300 °C. In contrast, the XRD pattern of the material obtained after 200 °C heat treatment showed no diffraction peaks, indicating the formation of an amorphous structure.

TEM verified the high crystallinity and phase purity of spinel $NiCo_2O_4$ NPs prepared by the low-temperature combustion method. Figure 2b displays a typical TEM image of the $NiCo_2O_4$ NPs synthesized at 250 °C. It can be seen that this material is composed of individual NPs with an average diameter of 4±1.3 nm, which is very close to the grain size calculated from XRD patterns. The high-resolution (HRTEM) image shown in Figure 2c reveals that the $NiCo_2O_4$ NPs possess a single-phase spinel structure with high crystallinity; combined with XRD results, the observed lattice fringes with interplanar distances ~2.4 Å and ~2.8 Å can be assigned to the (331) and (220) crystal planes of spinel $NiCo_2O_4$, respectively. The crystal structure of the $NiCo_2O_4$ NPs was further studied by selected-area electron diffraction (SAED). The SAED pattern taken from a small area of the $NiCo_2O_4$ NP aggregates (Figure 2d) shows a series of Debye-Scherrer diffraction rings, which can be assigned to the spinel phase of $NiCo_2O_4$. No other crystal phases were observed by means of electron diffraction. In addition, characterization of the chemical composition of $NiCo_2O_4$ NPs with energy-dispersive X-ray spectroscopy (EDS) showed an overall Ni:Co atomic ratio close to 1:2, in agreement with the stoichiometry of $NiCo_2O_4$ compound (**Figure S1**).

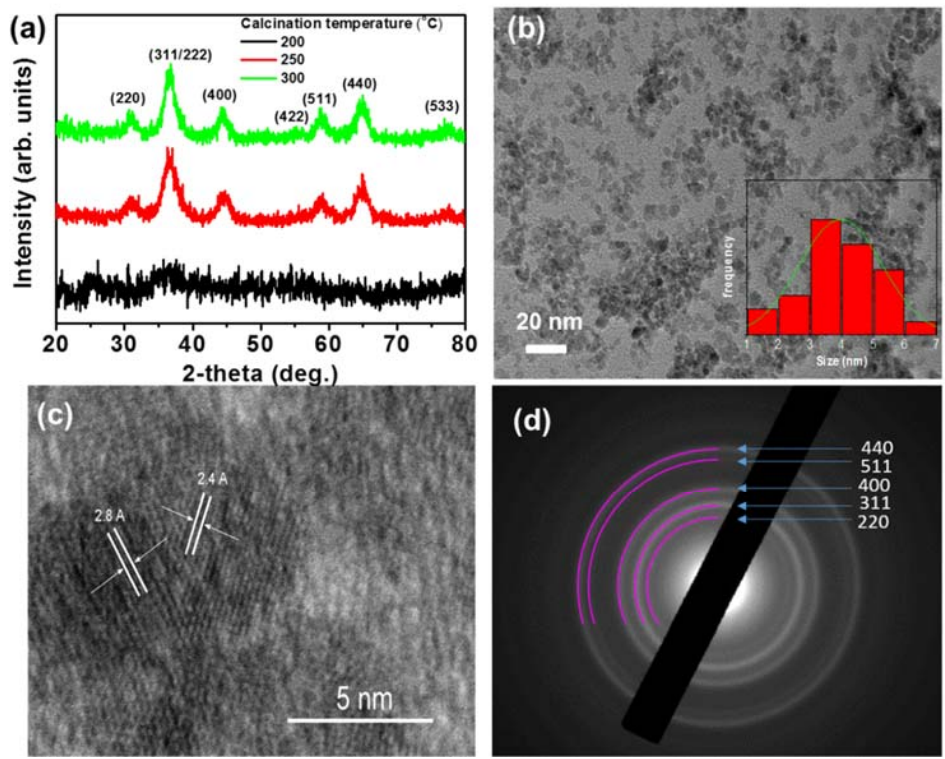

**Figure 2.** (a) XRD patterns of NiCo$_2$O$_4$ NPs at 200 ºC (black), 250 ºC (red), 300 ºC (green solid line) combustion temperatures. (b) Representative TEM image (inset: particle size distribution plot of the NiCo$_2$O$_4$ NPs at 250 ºC, showing an average diameter of 4 ± 1.3 nm), (c) high-resolution TEM and (d) SAED pattern of the as-synthesized NiCo$_2$O$_4$ NPs obtained at 250 ºC.

## 2.2. Blade Coating Processed Thin Films of NiCo$_2$O$_4$ NPs

Thin films of NiCo$_2$O$_4$ NPs were produced on top of quartz and ITO substrates using the blade coating technique, applying the processing parameters as described in the Experimental section. **Figure 3**, demonstrates the surface topography of a 15 nm-tick film of NiCo$_2$O$_4$ NPs fabricated on top of glass/ITO and quartz substrates, as obtained by AFM line scans. On top of ITO substrate (Figure 3a), the surface roughness is about

2.7 nm, while the film fabricated on quartz substrate (Figure 3b) exhibits an impressively smooth and compact topography of only 0.56 nm roughness. The development of a low roughness layer is a beneficial feature for the photovoltaic performance since it enables us to grow perovskite top layers with low roughness and enhanced homogeneity. Moreover, the dense $NiCo_2O_4$ NPs-based thin film exhibits an increase electrical conductivity up to 4 S/cm at room temperature, measured using four-point probe method.

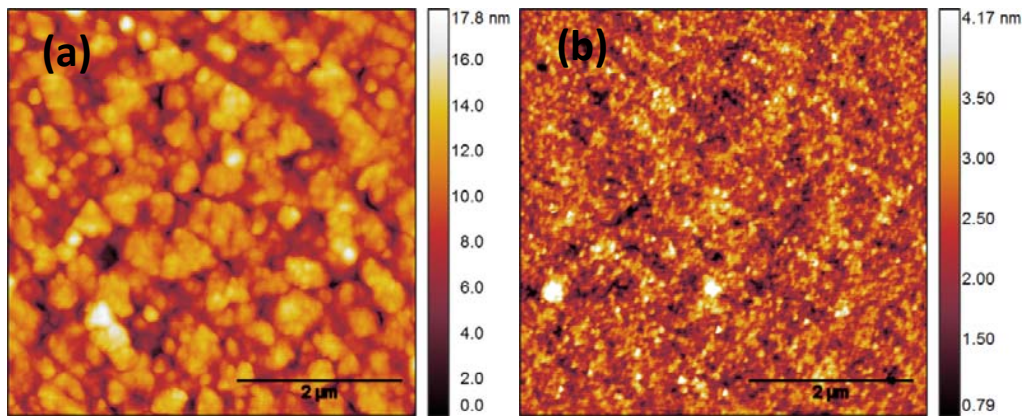

**Figure 3**. AFM images of (a) ITO/$NiCo_2O_4$ and (b) quartz/$NiCo_2O_4$ NPs thin films after combustion synthesis at 250 °C (The scale bar is 2 μm).

**Figure 4**a(inset), shows the optical absorption spectrum of the $NiCo_2O_4$ NPs film fabricated on a quartz substrate and the corresponding Tauc plot (Figure 4a) for direct allowed transition [$(\alpha E)^2$ *versus* photon energy (E)], giving an optical band gap of 2.32 eV. $NiCo_2O_4$ NPs thin films of different thicknesses were also fabricated on glass/ITO substrate in order to investigate the transmittance of the front contact at UV-vis spectrum. Figure 4b displays the transparency of bare glass/ITO and $NiCo_2O_4$ NPs coated on glass/ITO substrate; it could be seen that $NiCo_2O_4$ films thinner than 20 nm

reduce only slightly the transparency of the glass/ITO substrate for wavelengths longer than 450 nm, allowing intense light to reach the absorbing layer.

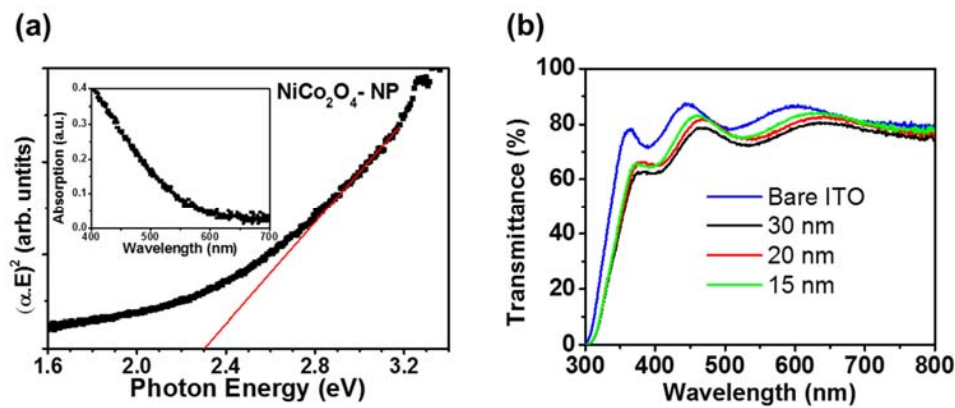

**Figure 4.** (α.E)$^2$ vs. photon energy (E) plot, showing an energy band gap of 2.32 eV. Inset: UV-vis absorption spectrum of NiCo$_2$O$_4$ NPs film fabricated on quartz substrate.(b) Transmittance spectra of bare glass/ITO and NiCo$_2$O$_4$ NPs films deposited on glass/ITO substrate with thickness of 15, 20 and 30 nm.

**2.3. Device Performance**

Complete p-i-n architectures of PVSCs were fabricated employing NiCo$_2$O$_4$ NPs thin films with three different thicknesses, 30, 20 and 15 nm. On top of each NiCo$_2$O$_4$ layer, a 230-nm-thick perovskite film was developed as described in Experimental section. The deposited perovskite film exhibits a low roughness of 5.4 nm (**Figure S2**) and a mean grain size of 0.22 μm with a standard deviation of 0.051 μm (**Figure S3**), as calculated by AFM topography measurements. To complete the devices, a PC[70]BM film was spun on each perovskite film serving as the electron transporting layer followed by a 100-nm-thick thermally deposited Al layer (**Figure 5**a). Figure 5c depicts the characteristic current density – voltage curves (J – V under 1 sun simulated intensity) of the PVSCs fabricated with 15, 20 and 30 nm-thick NiCo$_2$O$_4$ NPs layers, and the corresponding solar cell parameters are summarized in Table 1, where the series resistance (Rs) was extracted from the dark J-V curves (Figure 5d). It is observed that the J-V hysteric on the forward and reverse sweep is reduced as the thickness of $NiCo_2O_4$ decreases from 30 to 15 nm, while both the Voc and fill factor (FF) increase. Concretely, for the reversed sweep the Voc was increased from 0.90 to 0.99 V and the FF from 53.0 to 79.9 %, while the hysteric on the power conversion efficiency (PCE) for the 15-nm-thick $NiCo_2O_4$ layer is negligible. On the other hand, the short circuit current (Jsc) showed the lowest increase (~ 8%) for a forward sweep from 18.47 to 19.94 $mA/cm^2$, compared to both Voc and FF. Consequently, the device consisting of a 15-nm-thick $NiCo_2O_4$ film exhibits a PCE as high as 15.5 % for the forward sweep. The PCE of devices with thinner $NiCo_2O_4$ layers was declined (not shown here) exhibiting high leakage current due to not fully covered ITO. **Figure S4** demonstrates the external quantum efficiency (EQE) measurements of the corresponding devices. It is noticed that for 15-nm-thick $NiCo_2O_4$ film the overall efficiency is increased comparing to thicker layers due to higher transmittance as well as to better charge extraction, as it will be shown below. All the devices show a declined performance at longer wavelength (600 -750 nm) which can be ascribe to the relatively thin perovskite layer (~250 nm) and to not optimized back contact interface (Perovskite/PC[70]BM/Al).

The impact of the $NiCo_2O_4$ layer thickness on the ITO/$NiCo_2O_4$-NPs/perovskite device performance was evaluated by photoluminescence (PL) spectroscopy (see Figure 5b). Comparing to reference structure (without $NiCo_2O_4$ HTL), the PL signal of the devices with $NiCo_2O_4$ HTL show a quenching of more than 90 %, indicating a great reduction in the band-to-band charge recombination and, thus, a better hole selectivity of the ITO electrode covered by $NiCo_2O_4$. Further, the PL intensity is lower in 15 nm-

NiCo$_2$O$_4$ film than that of the thicker films, pointing to an efficient suppression of the electron-hole recombination (Figure 5b, inset).

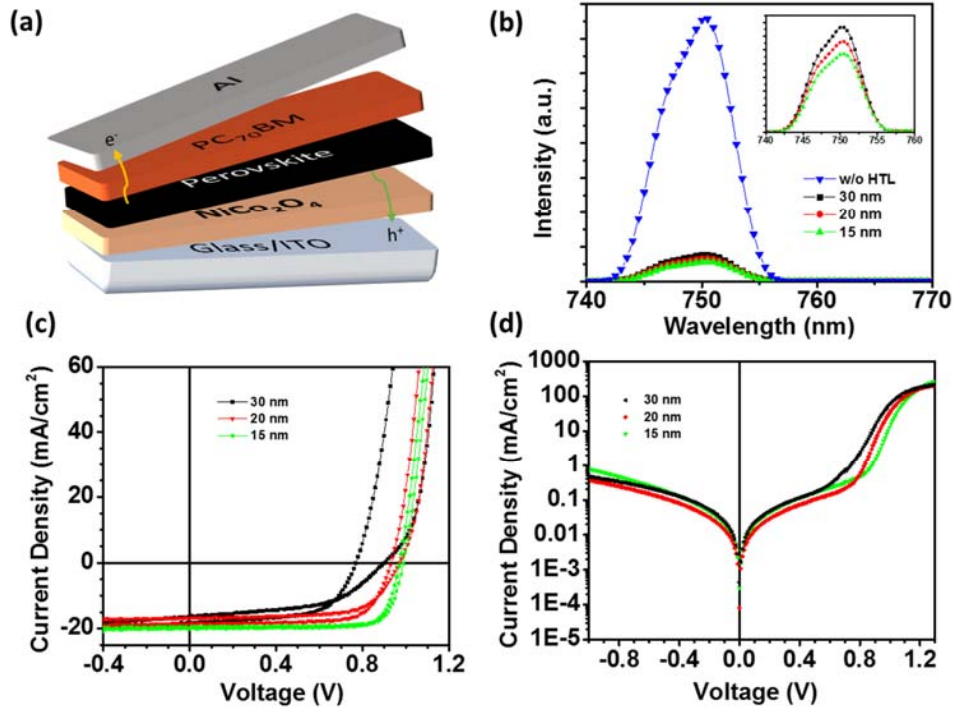

**Figure 5.** (a) The structure of the p-i-n perovskite solar cells under study (ITO/NiCo$_2$O$_4$-NPs/CH$_3$NH$_3$PbI$_3$/PC[70]BM/Al). (b) Photoluminescence (PL) spectra (inset: magnification of the PL spectra at lower intensities), and current density versus voltage (J-V) plots (c) under 1 sun illumination and (d) under dark conditions of the ITO/NiCo$_2$O$_4$-NPs/CH$_3$NH$_3$PbI$_3$ devices fabricated with NiCo$_2$O$_4$ with different thickness (15 nm - green solid line, 20 nm - red line and 30 nm – black line).

**Table 1.** Extracted solar cell parameters from the J – V characterization of the ITO/NiCo$_2$O$_4$/CH$_3$NH$_3$PbI$_3$/PC[70]BM/Al devices using NiCo$_2$O$_4$ NPs layers with different thickness.

| NiCo$_2$O$_4$ | Jsc [mA/cm$^2$] | Voc [V] | FF [%] | PCE [%] | Rs [$\Omega$cm$^2$] |
|---|---|---|---|---|---|
| 15 nm (forw.) | 19.94 | 0.97 | 79.9 | 15.5 | 1.06 |
| (rev.) | 19.60 | 0.99 | 79.2 | 15.4 | |
| 20 nm (forw.) | 18.47 | 0.93 | 73.2 | 12.6 | 1.34 |
| (rev.) | 16.83 | 0.97 | 67.8 | 11.1 | |
| 30 nm (forw.) | 18.45 | 0.77 | 61.2 | 8.7 | 1.37 |
| (rev.) | 16.29 | 0.90 | 53.0 | 7.8 | |

To further understand the charge recombination processes during the hole transfer from perovskite to NiCo$_2$O$_4$ layer, we performed electrochemical impedance spectroscopy (EIS) measurements under solar light and zero bias. **Figure 6**a shows the characteristic Nyquist plots of the three corresponding devices for 15, 20 and 30 nm-sized NiCo$_2$O$_4$ films. The results showed a shape of two frequency responses for PVSCs, where the second semicircle (feature at low frequencies) is been attributed to the recombination resistance (Rrec).[74,75] As the NiCo$_2$O$_4$ film thickness is reduced the radius of the semicircle increases, which implies a higher resistance in the charge recombination, in agreement with the findings from PL measurements. Figure 6b presents the Mott-Schottky (M-S) plots of the devices when sweeping from higher to a lower voltage. The crossing of the curves at 1/C$^2$ = 0 is attributed to the flat band potential of the device, while the lower slope of the linear region is ascribed to the charge accumulation at the interfaces, which impedes an efficient extraction of the charge carriers.[76] The M-S slope for the thicker film is lower implying that this layer cannot extract fast enough the charge carriers, inducing their accumulation at the interface. This behavior causes a higher hysteresis, which in turn increase electron-hole recombination (due to high spatial density) and leads to the drop of the flat band potential. On contrary, the thinner layer seems to extract faster the charge carriers, increasing the flat band potential, and thus the Voc, as well as the FF of the corresponding device. The direct correlation of faster charge carriers extraction at thinner HTL layers with increased FF has been previously studied by M.Stolterfoht *et. al.*.[77] We also notice that the configuration of 15 nm $NiCo_2O_4$ thin layer is depleted faster than the other devices due to enhanced charge carrier extraction, which can also affirm the increase at the FF of the corresponding PVSC . The increased Jsc for the 15 nm $NiCo_2O_4$ layer can be ascribed to the higher transparency of the film (shown above) and thus higher photogeneration of electron-hole pairs as well as to the lower series resistance as shown in Table 1.

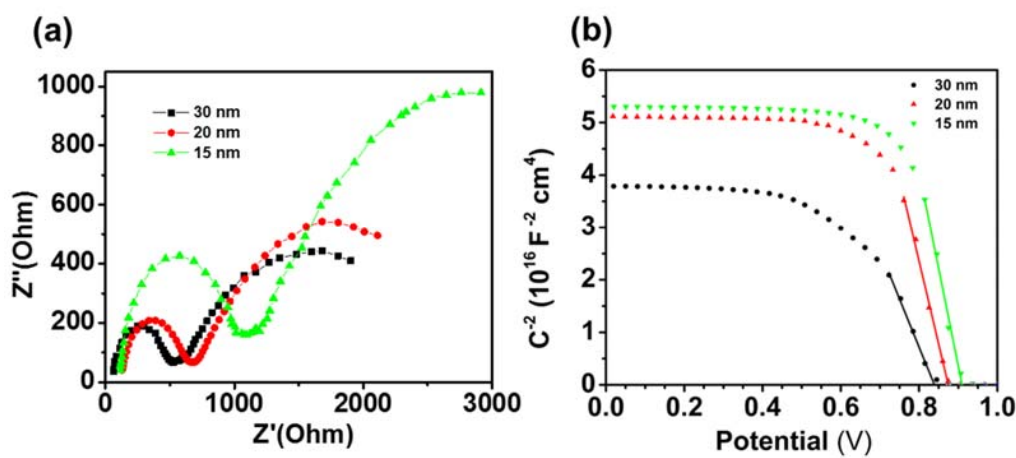

**Figure 6.** (a) Nyquist and (b) Mott-Schottky plots for the PVSK devices with 15, 20 and 30 nm thickness of the $NiCo_2O_4$ HTL.

Importantly, improved performance reproducibility and reliability of the corresponding p-i-n perovskite solar cells, compare to our previously reported solvothermal synthesized CuO-HTL based perovskite solar cells [22] was demonstrated by applying combustion synthesized $NiCo_2O_4$ NPs as HTL. Specifically, as shown in **Figure S5** the combustion synthesized $NiCo_2O_4$-HTL delivers a ~14.5 % average PCE (16 devices) with 15.5 % best performing device, while CuO-HTL based solar cells

give a ~12.5 % average PCE (16 devices) with 15.3 % best performing device.[22]

3. Conclusions

In conclusion, a low temperature combustion synthesis method, using for the first time a tartaric acid as a fuel, was successfully developed and applied for the fabrication of compact films of p-type NiCo2O4 NPs. The size of the NPs was fully controlled due to the usage of tartaric acid leading to the formation of monodispersed NiCo2O4 NPs with a diameter of ~4 nm. The combustion proceeds under low temperature (250 oC) and within a short reaction time (1 h), produce high quality, homogeneous NiCo2O4 NPs films with high electrical conductivity (~4 S.cm-1) and very low roughness (0.56 nm) functional layers were fabricated. The detailed physicochemical characterization of the NiCo2O4 NPs using X-ray diffraction, EDS and electron microscopy measurements confirm the high purity, crystallinity and small grain composition of the NiCo2O4. Furthermore, the proposed synthetic approach allowed the production of compact films using blade coating, which is a large-scale compatible technique appropriate for the development of printed electronic devices. The impact of NiCo2O4 HTL thicknesses on PVSCs characteristics were also investigated. The optimum thickness found to be 15 nm showing enhanced charge carrier collection and negligible J-V hysterics, compared to thicker films, delivering reliable p-i-n PVSCs with a PCE of 15.5 %. We believe that the proposed combustion synthesis method using a tartaric acid as a fuel can provide a route to produce highly reproducible metal oxides suitable for use in a range of advanced materials applications.

4. Experimental Section

*Materials:* Pre-patterned glass-ITO substrates (sheet resistance 4Ω/sq) were purchased from Psiotec Ltd, $Pb(CH_3CO_2)_2 \cdot 3H_2O$ from Alfa Aesar, MAI and MABr from Dyenamo Ltd, PC[70]BM from Solenne BV. All the other chemicals used in this study were purchased from sigma Aldrich.

*Synthesis of $NiCo_2O_4$ NPs films:* For the combustion synthesis of $NiCo_2O_4$ NPs, 0.5 mmol $Ni(NO_3)_2 \cdot 6H_2O$, 1 mmol $Co(NO_3)_2 \cdot 6H_2O$ and tartaric acid were mixed in the 15 ml 2-methoxy ethanol solution. After 150 uL $HNO_3$ (69% wt $HNO_3$) were added slowly into the mixture, and the solution stirred up to almost complete homogeneity. The whole solution were allowed under stirring for 30 min at 60 °C. The ratio of the total metal nitrates and tartaric acid was 1. Thereafter, the violet colored solution were used for the combustion synthesis of the $NiCo_2O_4$ nanoparticles on the various substrates. Doctor blade technique were applied for the fabrication of the precursor films on the various substrates. The resulting light violet colored films were dried at 100 °C for 30 min, and used as a precursor for the combustion synthesis of $NiCo_2O_4$ NPs. Subsequently the obtained films were heated at different temperatures (200 °C, 250 °C and 300 °C) in ambient atmosphere for 1 h in a preheated oven to complete the combustion process and then left to cool down at room temperature. For UV-Vis absorption measurements the films were fabricated on quartz substrates, while for the transmittance measurements 30, 20 and 15 nm thick films were fabricated on glass/ITO substrates applying 250 °C heating temperature respectively.

*Device fabrication:* The inverted solar cells under study was ITO/$NiCo_2O_4$-NPs/$CH_3NH_3PbI_3$/PC[70]BM/Al. ITO substrates were sonicated in acetone and subsequently in isopropanol for 10 min and heated at 100 °C on a hot plate 10 min before use. The perovskite solution was prepared 30 min prior spin coating by mixing

Pb(CH$_3$CO$_2$)$_2$.3H$_2$O:Methylamonium iodide (1:3) at 36 %wt in DMF with the addition of 1.5 % mole of methylamonium bromide (MABr).[78–80] The precursor was filtered with 0.1 µm PTFE filters. The perovskite precursor solution was deposited on the HTLs by static spin coating at 4000 rpm for 60 seconds and annealed for 5 minutes at 85 °C, resulting in a film with a thickness of ~230 nm. The PC[70]BM solution, 20 mg ml$^{-1}$ in chlorobenzene, was dynamically spin coated on the perovskite layer at 1000 rpm for 30 sec. Finally, 100 nm Al layers were thermally evaporated through a shadow mask to finalize the devices giving an active area of 0.9 mm$^2$. Encapsulation was applied directly after evaporation in the glove box using a glass coverslip and an Ossila E131 encapsulation epoxy resin activated by 365 nm UV-irradiation.

*Characterization:* Thermogravimetric (TGA) and Differential Thermal Analysis (DTA) were performed on a Shimadzu Simultaneous DTA-TG system (DTG-60H). Thermal analysis was conducted from 40 to 600 °C in air atmosphere (200 mL min$^{-1}$ flow rate) with a heating rate of 10 °C min$^{-1}$. X-ray diffraction (XRD) patterns were collected on a PANanalytical X´pert Pro MPD powder diffractometer (40 kV, 45 mA) using Cu Kα radiation (λ=1.5418 Å). Transmission electron microscope (TEM) images and electron diffraction patterns were recorded on a JEOL JEM-2100 microscope with an acceleration voltage of 200 kV. The samples were first gently ground, suspended in ethanol and then picked up on a carbon-coated Cu grid. Quantitative microprobe analyses were performed on a JEOL JSM-6390LV scanning electron microscope (SEM) equipped with an Oxford INCA PentaFET-x3 energy dispersive X-ray spectroscopy (EDS) detector. Data acquisition was performed with an accelerating voltage of 20 kV and 60 s accumulation time. Transmittance and absorption measurements were performed with a Schimadzu UV-2700 UV-Vis spectrophotometer. The thickness of the films were measured with a Veeco Dektak 150 profilometer. The current density-voltage (J/V) characteristics were characterized with a Botest LIV Functionality Test System. Both forward and reverse scans were measured with 10 mV voltage steps and 40 msec of delay time. For illumination, a calibrated Newport Solar simulator equipped with a Xe lamp was used, providing an AM1.5G spectrum at 100 mW/cm$^2$ as measured by a certified oriel 91150V calibration cell. A shadow mask was attached to each device prior to measurements to accurately define the corresponding device area. EQE measurements were performed by Newport System, Model 70356_70316NS. Atomic force microscopy (AFM) images were obtained using a Nanosurf easy scan 2 controller under the tapping mode. Electrical conductivity measurements were performed using a four-point microposition probe, Jandel MODEL RM3000. Electro-impedance spectroscopy (EIS) and Mott-Schottky measurements were performed using a Metrohm Autolab PGSTAT 302N, where for the EIS a red light-emitting diode (LED) (at 625 nm) was used as the light source calibrated to 100 mW/cm$^2$. For EIS a small AC perturbation of 20 mV was applied to the devices, and the different current output was measured throughout a frequency range of 1 MHz-1 Hz. The steady state DC bias was kept at 0 V throughout the EIS experiments.

**Supporting Information**

Supporting Information is available from the Wiley Online Library or from the author.

**Acknowledgements**

This project has received funding from the European Research Council (ERC) under the European Union's Horizon 2020 research and innovation programme (grant agreement No 647311).

# Supporting information

## Low Temperature Combustion Synthesis of a Spinel NiCo$_2$O$_4$ Hole Transport Layer for Perovskite Photovoltaics

*Ioannis T. Papadas,[a] Apostolos Ioakeimidis,[a] Gerasimos S. Armatas[b] and Stelios A. Choulis[a]\**

[a] Molecular Electronics and Photonics Research Unit, Department of Mechanical Engineering and Materials Science and Engineering, Cyprus University of Technology, Limassol, Cyprus.

[b] Department of Materials Science and Technology, University of Crete, Heraklion 71003, Greece.

\*Corresponding Author: Prof. Stelios A. Choulis

E-mail: stelios.choulis@cut.ac.cy

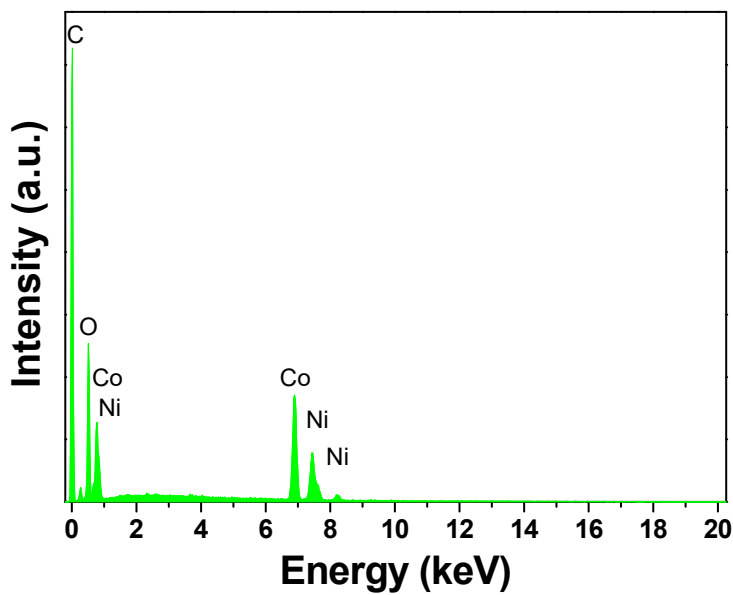

**Figure S1.** Typical EDS spectrum for NiCo$_2$O$_4$ nanoparticles. The EDS analysis indicates an average atomic proportion of Ni:Co ~ 1:2.

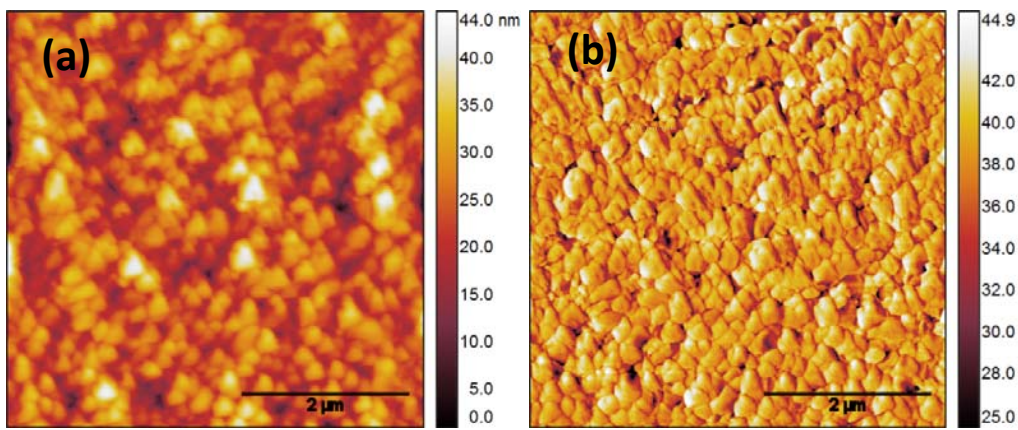

**Figure S2. (a)** Surface topography and **(b)** the corresponding phase image (5x5 μm) of 230 nm Perovskite film obtained by AFM. The film exhibit a roughness (root mean square) of 5.4 nm.

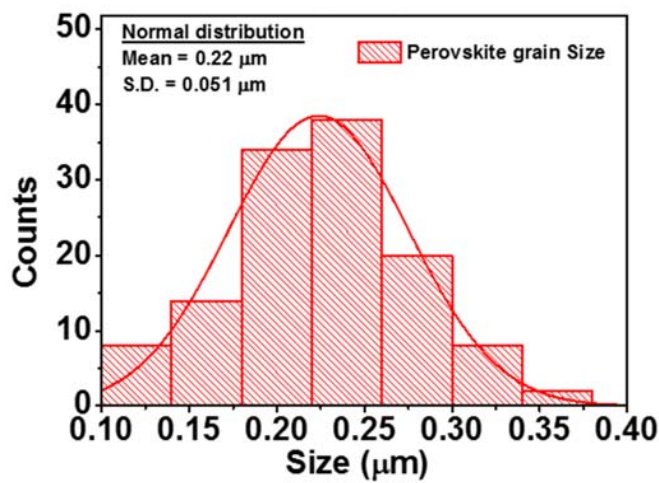

**Figure S3.** Size distribution of perovskite grains extracted from the AFM topography images. The mean size of the grains is 0.22 μm with a standard deviation of 0.051 μm.

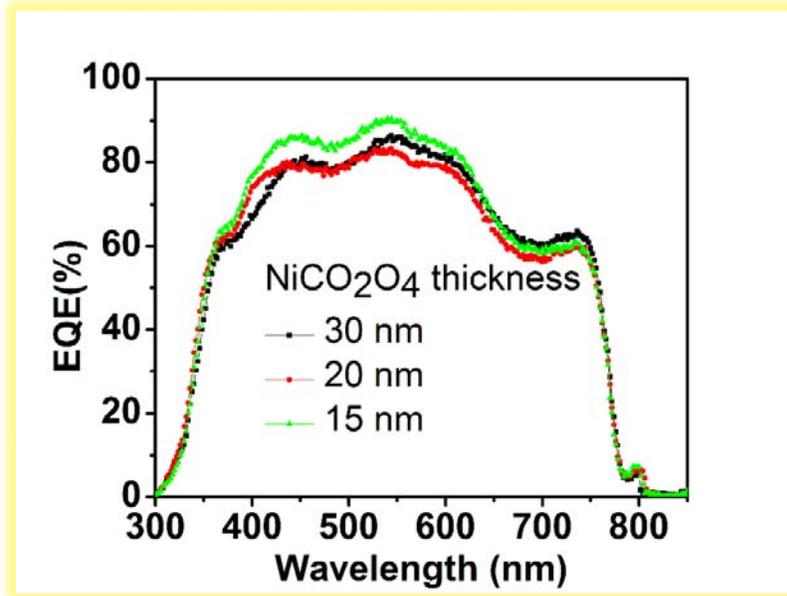

**Figure S4.** External Quantum Efficiency (EQE) of ITO/NiCo$_2$O$_4$-NPs/CH$_3$NH$_3$PbI$_3$ devices fabricated with NiCo$_2$O$_4$ with different thickness (15 nm - green solid line, 20 nm - red line and 30 nm – black line).

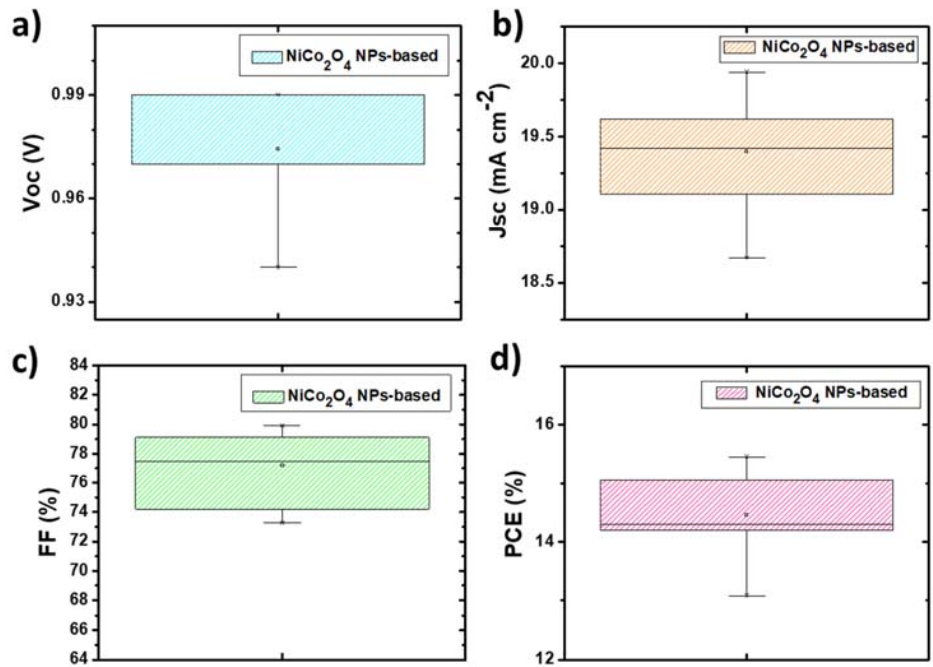

**Figure S5.** Average photovoltaic parameters represented in box plots out of 16 devices of each series of p-i-n perovskite solar cells under study. NiCo$_2$O$_4$ NPs-based devices with box plots, a) open circuit voltage (Voc), b) current density (Jsc), c) fill factor (FF) and d) power conversion efficiency (PCE).